\documentstyle[manuscript,aps]{revtex}
\begin{document}
\preprint{
\begin{tabular}{r}
MIT-CTP-2735\ \
\end{tabular}
} 
\begin{titlepage}
\title{An Explicit Model of Quark Mass Matrix}
\author{Elisabetta Sassaroli \thanks{This work is supported in part by funds 
provided by the U.S. Department of Energy (D.~O.~E.) under 
cooperative research agreement $\#$DF-FC02-94ER40818}  }
\address{Laboratory for Nuclear Science and Department of Physics}
\address{Massachusetts Institute of Technology, Cambridge, MA 02139}
\address{Physics Department, Northeastern University, Boston, MA 02115} 
\maketitle

\begin{abstract}
A physical model which describes the CKM matrix is analyzed. 
The elements of such a matrix are field-strength renormalization factors. 
Each column gives  the probability amplitude for the 
field operators of the coupled Lagrangian to create a one-particle 
eigenfunction of definite energy. The total conserved charge is the sum of 
the flavor charges which are not conserved separately.

\end {abstract}

\thispagestyle{empty}

\end{titlepage}

\section{Introduction}

Although the Standard Model is a successful theory, its description of quark
masses and their flavor mixing is uncertain. It is generally believed that
physics beyond the Standard Model is necessary in order to acquire a
complete understanding of the quark mass problem. In the absence of a
satisfactory theory for describing the phenomenon of flavor mixing, a number
of different parameterizations of the Cabibbo-Kobayashi-Maskawa matrix (CKM) 
\cite{Ca,Ko}, have been considered in the literature, see for example Refs.  
\cite{Ma,Ch,Di,Fr,Ra,Ba,Ras} and references there. For a detailed review of
the different parameterizations see Ref. \cite{Fri}. 

It is hoped that once the parameters of the CKM matrix will be known more
precisely, they can give us some clue about the physics beyond the Standard
Model. For example a given parameterization may be  more useful than
others in analyzing a given structure of a quark mass matrix. This because it
may suggest different kinds of modifications when one tries to understand
what is not yet understood.

In this paper an explicit model of the CKM  rotation matrix  is
considered. This represents a somewhat different approach with respect to the
previous ones and it should be considered as complementary to them. This
model is obtained by the diagonalization of the 
Lagrangian 
$$
{\cal L}={\bar \psi }_{d^{\prime }}(i\gamma \cdot \partial -m_{d^{\prime
}})\psi _{d^{\prime }}+{\bar \psi }_{s^{\prime }}(i\gamma \cdot \partial
-m_{s^{\prime }})\psi _{s^{\prime }}+{\bar \psi }_{b^{\prime }}(i\gamma
\cdot \partial -m_{b^{\prime }})\psi _{b^{\prime }} 
$$
$$
-({\delta \bar \psi }_{d^{\prime }}\psi _{s^{\prime }}+\delta ^{*}{\bar \psi 
}_{s^{\prime }}\psi _{d^{\prime }})-(\varepsilon {\bar \psi }_{d^{\prime
}}\psi _{b^{\prime }}+\varepsilon ^{*}{\bar \psi }_{b^{\prime }}\psi
_{d^{\prime }})-({\eta \bar \psi }_{s^{\prime }}\psi _{b^{\prime }}+\eta ^{*}%
{\bar \psi }_{b^{\prime }}\psi _{s^{\prime }}),\eqno(1.1) 
$$
where $\psi _{d^{\prime }}$, $\psi _{s^{\prime }}$ and $\psi _{b^{\prime }}$
are quark flavor fields with masses respectively $m_{d^{\prime }}$, $%
m_{s^{\prime }}$, and $m_{b^{\prime }}$. The complex coupling constants $%
\delta $, $\varepsilon $ and $\eta $ describe flavor mixing.

The Lagrangian given by Eq. (1.1) describes quarks which are in a state of
mixed flavors at any space time point. The rotation matrix is obtained by
the diagonalization of the above Lagrangian. The elements of such a matrix are
field strength renormalization factors and are functions of the quark masses 
$m_{d^{\prime }}$, $m_{s^{\prime }}$ and $m_{b^{\prime }}$ and their
coupling constants. This point is certainly not obvious in the previous
treatments and it gives some physical insight into the meaning of the CKM rotation
matrix. 

\section{The Quark Mass Matrix}

\subsubsection{Energy Eigenvalues}

In order to diagonalize the Lagrangian given by Eq. (1.1) the procedure
illustrated in ref. \cite{Sa} is generalized to the three flavor case. For this
case the following system of three coupled equations is diagonalized 
$$
\begin{array}{c}
i 
\frac{\partial \psi _{d^{\prime }}}{\partial t}=(\alpha \cdot 
\overrightarrow{p}+\beta m_{d^{\prime }})\psi _{d^{\prime }}+\beta \delta
\psi _{s^{\prime }}+\beta \varepsilon \psi _{b^{\prime }}, \\ i 
\frac{\partial \psi _{s^{\prime }}}{\partial t}=(\alpha \cdot 
\overrightarrow{p}+\beta m_{s^{\prime }})\psi _{s^{\prime }}+\beta \delta
^{*}\psi _{d^{\prime }}+\beta \eta \psi _{b^{\prime }}, \\ i\frac{\partial
\psi _{b^{\prime }}}{\partial t}=(\alpha \cdot \overrightarrow{p}+\beta
m_{b^{\prime }})\psi _{b^{\prime }}+\beta \varepsilon ^{*}\psi _{d^{\prime
}}+\beta \eta ^{*}\psi _{s^{\prime }}, 
\end{array}
\eqno(2.1) 
$$
where $\alpha $ and $\beta $ are Dirac matrices. We notice here that the
Hamiltonian is hermitian, therefore the eigenvalues are real. The energy
eigenvalues are obtained by solving the cubic equation 
$$
z^3-z^2(3p^2+h)+z(3p^4+2p^2h+l)-p^6-p^4h-p^2l-f=0,\eqno(2.2) 
$$
where $z=E^2,$ and the parameters {\it h, f, l} are given by 
$$
\begin{array}{c}
h=m_{d^{\prime }}^2+m_{s^{\prime }}^2+m_{b^{\prime }}^2+2(|\delta
|^2+|\varepsilon |^2+|\eta |^2), \\ 
f=\left[ r+m_{d^{\prime }}m_{s^{\prime }}m_{b^{\prime }}-(m_{d^{\prime
}}|\eta |^2+m_{s^{\prime }}|\varepsilon |^2+m_{b^{\prime }}|\delta
|^2)\right] ^2, \\ 
r=\varepsilon \delta ^{*}\eta ^{*}+\varepsilon ^{*}\eta \delta , \\ 
l=l_1+l_2, \\ 
l_1=(|\delta |^2-m_{d^{\prime }}m_{s^{\prime }})^2+(|\varepsilon
|^2-m_{d^{\prime }}m_{b^{\prime }})^2+(|\eta |^2-m_{s^{\prime }}m_{b^{\prime
}})^2, \\ 
l_2=2\left[ |\eta |^2(|\varepsilon |^2+m_{d^{\prime }}^2)+|\varepsilon
|^2(|\delta |^2+m_{s^{\prime }}^2)+|\delta |^2(|\eta |^2+m_{b^{\prime
}}^2)-r(m_{d^{\prime }}+m_{s^{\prime }}+m_{b^{\prime }})\right] . 
\end{array}
\eqno(2.3) 
$$
Eq.(2.1) can be further simplified by the transformation%
$$
z=E^2=x+\frac{3p^2+h}3,\eqno(2.4) 
$$
which brings Eq. (2.1) to the simpler form 
$$
x^3+x(l-\frac{h^2}3)+\frac{hl}3-\frac{2h^3}{27}-f=0.\eqno(2.5) 
$$
If we assume%
$$
\begin{array}{c}
a^3+b^3= 
\frac{hl}3-\frac{2h^3}{27}-f, \\ ab=-\frac 13(l-\frac{h^2}3), 
\end{array}
\eqno(2.6) 
$$
the Eq. (2.5) becomes%
$$
x^3-3abx+a^3+b^3=0,\eqno(2.7) 
$$
which has the roots 
$$
-a-b,\ \ -\omega a-\omega ^2b,\ \ -\omega ^2a-\omega b,\eqno(2.8) 
$$
where $\omega $ is one of the complex cube roots of unity, i.e. $\omega
^3=1. $

Then $a^3$ and {\it b}$^3$ are the roots of the quadratic equation%
$$
y^2-y(\frac{hl}3-\frac{2h^3}{27}-f)-\frac 1{27}(l-\frac{h^2}3)^3=0,\eqno(2.9)
$$
with solutions%
$$
y_{1,2}=\frac 1{54}\left( -27f-2h^3+9hl\pm 3\sqrt{3}\sqrt{%
27f^2+4fh^3-18fhl-h^2l^2+4l^3}\right) .\eqno(2.10) 
$$

We are looking for real eigenvalues, hence the discriminant in Eq.
(2.10) must be negative. Therefore $a^3$ and $b^3$ are complex numbers, and
in general they have to be found by using  De Moivre's theorem. If we
assume that $a=m+in$ and $b=m-in$ with 
$$
\begin{array}{c}
m^3-3mn^2= 
\frac 1{54}(-27f-2h^3+9hl), \\ 3nm^2-n^3=\frac 1{54}\left( 3\sqrt{3}\sqrt{%
|27f^2+4fh^3-18fhl-h^2l^2+4l^3|}\right) , 
\end{array}
\eqno(2.11) 
$$
then the real roots of Eq. (2.5) are 
$$
-2m,\ \ m+n\sqrt{3},\ \ m-n\sqrt{3}.\eqno(2.12) 
$$
Thus to find the three real roots of Eq. (2.5) we have to find the cube
roots of two complex numbers. For practical purposes it is better to find
one real root $x_1$ by a numerical method and then divide out the factor $%
x-x_1$ to obtain a quadratic equation.

The three energy eigenvalues can be formally written as 
$$
E_{1,2,3}^2=p^2+m_{1,2,3}^2,\eqno(2.13) 
$$
where the renormalized masses $m_{1,2,3}^2$ are given respectively by 
$$
\begin{array}{c}
m_1^2= 
\frac h3-2m, \\ m_2^2= 
\frac h3+m+n\sqrt{3}, \\ m_3^2=\frac h3+m-n\sqrt{3}. 
\end{array}
\eqno(2.14) 
$$

\subsubsection{Field-Strength Renormalization Constants}

The energy eigenfunctions associated with the eigenvalues $E_{1,2,3}$ are of
the type%
$$
\begin{array}{c}
\psi _i=N_i\left( 
\begin{array}{c}
M_{i1} \\ 
M_{i2} \\ 
M_{i3} 
\end{array}
\right) 
\frac 1{\sqrt{V}}\frac{u_i(s,{\bf p})}{\sqrt{2E_i}}e^{-E_it}e^{i{\bf p}\cdot 
{\bf x}}, \\  
\end{array}
\eqno(2.15) 
$$
where $u_i$ is the Dirac spinor of mass $m_i$ and the coefficients $%
M_{i1,2,3}$ are given by%
$$
\begin{array}{c}
M_{i1}=1, \\ 
M_{i2=} 
\frac{C-m_i^2D}{A-m_i^2B}, \\ M_{i3}=\frac{G+m_i^2F-m_i^4}{A-m_i^2B}, 
\end{array}
\eqno(2.16) 
$$
with

$$
\begin{array}{c}
A=\delta \eta (|\delta |^2+|\eta |^2-m_{d^{\prime }}m_{s^{\prime
}}-m_{d^{\prime }}m_{b^{\prime }}-m_{s^{\prime }}m_{b^{\prime
}})+\varepsilon [m_{s^{\prime }}^2(m_{d^{\prime }}+m_{b^{\prime }})+ \\ 
|\delta |^2(m_{b^{\prime }}-m_{s^{\prime }})-|\eta |^2(m_{s^{\prime
}}-m_{d^{\prime }})]-\varepsilon ^2\delta ^{*}\eta ^{*}, \\ 
B=\delta \eta +\varepsilon (m_{d^{\prime }}+m_{s^{\prime }}), \\ 
C=\eta [m_{d^{\prime }}^2(m_{s^{\prime }}+m_{b^{\prime }})+|\delta
|^2(m_{b^{\prime }}-m_{d^{\prime }})+|\varepsilon |^2(m_{s^{\prime
}}-m_{d^{\prime }})) \\ 
-\varepsilon \delta ^{*}(m_{d^{\prime }}m_{s^{\prime }}+m_{d^{\prime
}}m_{b^{\prime }}+m_{s^{\prime }}m_{b^{\prime }}-|\delta |^2-|\varepsilon
|^2)-\delta \eta ^2\varepsilon ^{*}] \\ 
D=\varepsilon \delta ^{*}+\eta (m_{s^{\prime }}+m_{b^{\prime }}) \\ 
G=(m_{d^{\prime }}+m_{s^{\prime }})(\delta \eta \varepsilon ^{*}+\varepsilon
\delta ^{*}\eta ^{*})-(m_{d^{\prime }}m_{s^{\prime }}-|\delta
|^2)^2-m_{s^{\prime }}^2|\varepsilon |^2-m_{d^{\prime }}^2|\eta |^2- \\ 
|\delta |^2|\varepsilon |^2-|\delta |^2|\eta |^2 \\ 
F=m_{d^{\prime }}^2+m_{s^{\prime }}^2+2|\delta |^2+|\varepsilon |^2+|\eta
|^2. 
\end{array}
\eqno(2.17) 
$$
$N_i$ is the normalization constant given by%
$$
N_i=\frac{|A-m_i^2B|}{\sqrt{|A-m_i^2B|^2+|C-m_i^2D|^2+|G+m_i^2F-m_i^4|^2}}.%
\eqno(2.18) 
$$

The eigenfunctions $\psi _{1,2,3}$ represent states of given energy but of
mixed flavors at any space time point. As discussed in ref.\cite{Sa} in field theory the
elements of the $3\;\times 1$ matrix in Eq. (2.15) are field-strength
renormalization factors and give the probability amplitude for the field
operators $\psi_{d',s',b'}$ to create one-particle eigenfunctions of energy $%
E_{1.2,3}$ and mixed flavors.

The rotation matrix between the flavor base and the mass base is given by
these field-strength renormalization factors and it can be written as%
$$
M=\left( 
\begin{array}{ccc}
N_1 & N_2 & N_3 \\ 
N_1\frac{C-m_1^2D}{A-m_1^2B} & N_2\frac{C-m_2^2D}{A-m_2^2B} & N_3 
\frac{C-m_3^2D}{A-m_3^2B} \\ N_1\frac{G+m_1^2F-m_i^4}{A-m_1^2B} & N_2\frac{%
G+m_2^2F-m_2^4}{A-m_2^2B} & N_3\frac{G+m_3^2F-m_3^4}{A-m_3^2B} 
\end{array}
\right) .\eqno(2.19) 
$$
In the limit of only two flavor the rotation matrix reduces to%
$$
U=\left( 
\begin{array}{cc}
\frac 1{\sqrt{1+M_1^2}} & \frac 1{\sqrt{1+M_2^2}} \\ \frac{M_1}{\sqrt{1+M_1^2%
}} & \frac{M_2}{\sqrt{1+M_2^2}} 
\end{array}
\right) =\left( 
\begin{array}{cc}
\frac 1{\sqrt{1+M_1^2}} & \frac{M_1}{\sqrt{1+M_1^2}} \\ \frac{M_1}{\sqrt{%
1+M_1^2}} & -\frac 1{\sqrt{1+M_1^2}} 
\end{array}
\right),\eqno(2.20) 
$$
where $M_{1\text{ }}$is given by%
$$
M_1=\frac{m_{d^{\prime }}-m_{s^{\prime }}+\sqrt{(m_{d^{\prime
}}-m_{s^{\prime }})^2+4|\delta |^2}}\delta =-\frac 1{M_2}.\eqno(2.21) 
$$
Therefore the matrix $U$ is equivalent to the 
Cabibbo matrix.

Even if not manifestly, the matrix M is unitary, this 
because the wavefunctions given by Eq. (2.15) are orthonormal. 
The possibility of a quark rotation matrix without the
assumption of unitarity has been discussed in \cite{Fra,Sw}

The matrix $M$ given by Eq. (2.17) also can describe neutrino flavor mixing,
however only in the ultra-relativistic limit for the reasons discussed in 
\cite{Sa} .

\section{Summary}

We have derived a physical representation of the CKM matrix. The conserved
current is the sum of the three flavor currents and the elements of the
rotation matrix are field-strength renormalization constants. Because of its
simplicity it is hoped that this model can give some physical meaning to the
CKM matrix especially in relation to its improvement.

\end{document}